\documentclass[prl,twocolumn,superscriptaddress]{revtex4}
\usepackage{amssymb,amsmath,amsthm,graphicx}
\usepackage{latexsym}
\usepackage{bm}
\usepackage[]{hyperref}
\allowdisplaybreaks

\pdfoutput=1

\begin{document}

\title{Violating weak cosmic censorship in AdS$_4$}

\author{Toby Crisford}
\email{tc393@cam.ac.uk}
\affiliation{Department of Applied Mathematics and Theoretical Physics, University of Cambridge, Wilberforce Road, Cambridge CB3 0WA, UK} 
\author{Jorge E. Santos}
\email{jss55@cam.ac.uk}
\affiliation{Department of Applied Mathematics and Theoretical Physics, University of Cambridge, Wilberforce Road, Cambridge CB3 0WA, UK} 

\begin{abstract}
We consider time-dependent solutions of the Einstein-Maxwell equations using anti-de Sitter (AdS) boundary conditions, and provide the first counterexample to the weak cosmic censorship conjecture in four spacetime dimensions. Our counterexample is entirely formulated in the Poincar\'e patch of AdS. We claim that our results have important consequences for quantum gravity, most notably to the weak gravity conjecture.
\end{abstract}

\maketitle

{\bf~Introduction --} The weak cosmic censorship conjecture (WCCC) was originally formulated almost fifty years ago \cite{Penrose:1969pc}. Since then it has occupied a central role in cosmology, high energy physics, astrophysics and mathematics. There are many versions of the WCCC, and here we will be interested in whether or not it is possible to form a region of arbitrarily large curvature that is visible to distant observers. In higher dimensions, there is abundant numerical evidence that this is possible, so that this version of the WCCC does not hold \cite{Gregory:1993vy,Lehner:2010pn,Santos:2015iua,Figueras:2015hkb,Figueras:2017zwa}. However, in all of the previous scenarios, one starts with an unstable black hole solution, and it is far from clear whether such a black hole can be formed in the first place. The example we provide in this letter is four-dimensional and starts in the vacuum of the theory.

It is perhaps surprising that the WCCC plays an important role in high energy physics. However, through gauge-gravity duality, some gravitational configurations with asymptotically anti-de Sitter (AdS) boundary conditions can be mapped into states of strongly coupled field theories \cite{Maldacena:1997re,Gubser:1998bc,Witten:1998qj,Aharony:1999ti}. One can learn much about field theories by studying their gravitational dual, and vice-versa. For instance, we will argue that our counterexample to the WCCC may be connected to the weak gravity conjecture \cite{ArkaniHamed:2006dz}, which states that any consistent quantum theory of gravity must contain charged particles in its spectrum with a charge greater than or equal to their mass. It is remarkable that these two conjectures can be related at all \footnote{We thank C.~Vafa for pointing out this tantalising connection.}.

We restrict ourselves to solutions of the bulk Einstein-Maxwell action endowed with a negative cosmological constant
\begin{equation}
S=\frac{1}{16\pi G}\int_{\mathcal{M}}\mathrm{d}^4x\sqrt{-g}\left(R+\frac{6}{L^2}-F^2\right)\,,
\label{eq:action}
\end{equation}
where $L$ is the AdS curvature length scale and $F=\mathrm{d}A$ is the Maxwell field strength. Since AdS has a timelike conformal boundary, which we denote by $\partial \mathcal{M}$, we are free to specify the (conformal) boundary metric and the asymptotic form of the boundary Maxwell field strength $f\equiv \left.F\right|_{\partial \mathcal{M}}$. We fix the boundary metric to be conformally flat, and use polar coordinates $(R,\phi)$ to parametrise the boundary directions, that is to say, we take the conformal boundary metric to be given by
$$
\left.\mathrm{d}s^2\right|_{\partial \mathcal{M}}=-\mathrm{d}t^2+\mathrm{d}R^2+R^2\mathrm{d}\phi^2\,.
$$
For $f$ we choose a localised source that depends on $t$ and $R$, \emph{i.e.} $f=f(t,R)$. For simplicity we will restrict ourselves to axisymmetric configurations for which $\partial/\partial \phi$ is a Killing vector field.

In \cite{Horowitz:2014gva} static configurations with a boundary electric field profile of the form
\begin{equation}
\label{eq:profile}
f=\frac{a\,R\,\gamma}{\displaystyle\sigma^2\left(1+\frac{R^2}{\sigma^2}\right)^{\frac{\gamma}{2}+1}}\mathrm{d}t\wedge \mathrm{d}R\,,\quad\text{with}\quad \gamma\geq1
\end{equation}
were studied. Due to the conformal invariance of the UV theory, only the product $a\sigma$ is physically meaningful. From here onwards we set $\sigma = 1$. In all cases, a critical amplitude $a_{\max}$ was found, above which no static solution with a simply connected horizon could be constructed. For $\gamma>1$ and above some amplitude $a^{\star}<a_{\max}$, solutions with charged extremal spherically symmetric horizons hovering above the Poincar\'e horizon were found and seem to exist for arbitrarily large values of $a>a_{\star}$. For $\gamma=1,$ such hovering solutions have not yet been constructed.

We can now present our counterexample to the WCCC. We first promote the amplitude $a$ to be a function of time $a(t)$. Far in the past, we demand $a(-\infty)=0$ and we then increase $a(t)$ slowly over time to an amplitude $a(t)>a_{\max}$. Since the action (\ref{eq:action}) contains no charged matter, charged hovering black holes cannot form. This lead the authors in \cite{Horowitz:2016ezu} to conjecture that the endpoint of such a \emph{Gedankenexperiment} would ultimately provide a counterexample to the WCCC. This is precisely what we aim to address in this manuscript.

{\bf~Numerical method --} We solve the Einstein-Maxwell equations numerically using a characteristic scheme combined with spectral methods. We follow the procedure described in \cite{Balasubramanian:2013yqa} with two significant modifications: the inclusion of a Maxwell field, and a coordinate choice which is better adapted to describing an extremal horizon. The full details are given in supplemental material, along with checks on the accuracy of the numerics.

To apply a characteristic scheme it is first necessary to choose a set of null geodesics to use as the characteristic curves. This is equivalent to choosing coordinates corresponding to a null foliation of the space-time. In \cite{Balasubramanian:2013yqa} null geodesics are chosen orthogonal to constant $t$ slices of the conformal boundary. They can be labelled by their point of intersection with the boundary, $(t,R,\phi),$ and parametrised by a parameter, $z,$ to give coordinates for the bulk space-time. This works well at finite temperature, but if the horizon is extremal then it can be shown that all of these characteristics intersect the horizon at a single point. We would like to choose characteristics which cover both the entire conformal boundary and the entire horizon.

To achieve this, at points on the boundary with coordinates $(t, R),$ we pick a null geodesic which makes an angle $\tan{\psi} = c/R$ with the constant $t$ slice of the boundary, where $c$ is a constant. We parametrise the geodesic with a parameter, $\xi.$ As $R \rightarrow 0$ the null geodesics become orthogonal to the boundary, while as $R \rightarrow \infty$ they become parallel to it. It is helpful to think of this as a Cartesian to polar transformation on the bulk space-time sending $z$ and $R$ to $\xi$ and $\psi,$ where the origin of the polar coordinates lies outside the boundary. We find it convenient to use\begin{align}
\eta &= \sin{\psi} = \frac{c}{\sqrt{c^2 + R^2}}\\
v &= t - c/\eta
\end{align}as new boundary coordinates, and to label our characteristics with $(v,\eta,\phi)$ corresponding to their point of intersection with the boundary.

This method of constructing coordinates leads us to a metric of the form:
\begin{multline}
\mathrm{d}s^2 = \frac{L^2}{(\eta - c\,\xi)^2}\Big[-(\xi^2\,e^{2\beta} V - e^{2\chi + \alpha} U^2) \mathrm{d}v^2 + 2\,e^{2\beta} \mathrm{d}v\,\mathrm{d}\xi \\ - 2\,e^{2\chi+\alpha} U\,\mathrm{d}v\,\mathrm{d}\eta + e^{2\chi+\alpha} \mathrm{d}\eta^2 + e^{2\chi - \alpha}\mathrm{d}\phi^2\Big].
\end{multline}
We set $L$ and $c$ to $1$ in our numerics. There is remaining gauge freedom corresponding to redefinitions of $\xi,$ which can be used to fix $\chi.$ Following \cite{Balasubramanian:2013yqa} we use this freedom to impose that $\xi = 0$ is an apparent horizon. At time $v = 0,$ we will begin in pure AdS and $\xi = 0$ will be the Poincar\'e horizon. This allows us to ignore the $\xi < 0$ region of the space-time, and restrict to $0 \leq \xi < \eta/c.$ The conformal boundary lies at $\xi = \eta/c.$ Pure AdS space-time in these coordinates corresponds to $V=1,\ \beta=0,\ U=0,\ \alpha=-\log(1-\eta^2),\ \chi = 0.$ We use these as Dirichlet conditions at $v = 0$ and $\xi = \eta/c.$

There are three independent non-vanishing components of the Maxwell field strength tensor: $F_{v \eta},F_{v \xi},F_{\xi \eta}.$ As initial conditions, we take them all to vanish at $v = 0.$ As our boundary condition, we impose equation (\ref{eq:profile}) with $\gamma = 1$ and promote $a$ to be a function of $v = t - \sqrt{c^2+R^2}.$ In the new coordinates with $c = 1,$ this condition takes the remarkably simple form\begin{equation}
F_{v \eta} + F_{v \xi} = a(v)
\end{equation}
and we pick a time dependence for $v > 0:$
\begin{equation}
a(v) = a_{0} \left[1 - \text{sech}(5 v)\right].
\end{equation}
This function increases smoothly from an initial value of $0$ to a maximum value $a_0,$ with $a(1) > 0.98 a_{0}$ and $a(2) > 0.9999 a_{0}.$

Our choice of a simple $v$ dependence for the boundary electric field comes at the expense of introducing a more complicated $t$ and $R$ dependence. We nevertheless still have the important property that at large $t$ the electric field converges to the stationary profile (\ref{eq:profile}). We therefore expect to see a violation of the WCCC based on the arguments in \cite{Horowitz:2016ezu}.

{\bf~Results --} For the profile we consider, the critical amplitude was studied in \cite{Horowitz:2014gva} and found to be $a_{max} \approx 0.678.$ We have collected results up to time $v = 7.5$ for five amplitudes: $a_0 \approx 0.4243,\ a_0 \approx 0.5657,\ a_0 \approx 0.7071,\ a_0 \approx 0.8485,\ a_0 \approx 0.9899.$ The two solutions with sub-critical amplitude approach a smooth stationary solution at late times as expected. The three solutions with super-critical amplitude have curvature growing without bound on the event horizon at the point $\eta = 1.$ This growth appears to follow a power law in $v$, and so no singularity forms in finite time. However, by waiting for sufficiently long times we can form arbitrarily large curvatures visible to boundary observers, violating the WCCC \footnote{In critical gravitational collapse, given an arbitrarily large curvature bound it is possible to find an open set of initial data which will exceed it. The difference here is that we have generic initial data for which the curvature grows without \emph{any} bound.}.

To demonstrate this, we compute the value of $F^2 = F^{a b} F_{a b}$ on the apparent horizon ($\xi = 0$) at $\eta = 1$ and look at how it changes with $v.$ This is a gauge invariant scalar, so a divergence in $F^2$ indicates the formation of a singularity. We can also use the equations of motion to show that it implies a divergence in the space-time curvature as well. The Einstein-Maxwell equations imply that\begin{equation}
R_{a b} R^{a b} = \frac{36}{L^4}+4\left({F_{a}}^{c} F_{b c} F^{a d} {F^{b}}_{d}-\frac{1}{4} (F^2)^2\right)
\end{equation}
and we have checked that in our solutions\begin{equation}
{F_{a}}^{c} F_{b c} F^{a d} {F^{b}}_{d} = \frac{1}{2} (F^2)^2.\end{equation}A divergence in $F^2$ therefore implies a divergence in the curvature invariant $R_{a b} R^{a b}.$

\begin{figure}
\centering
\includegraphics[scale=0.3]{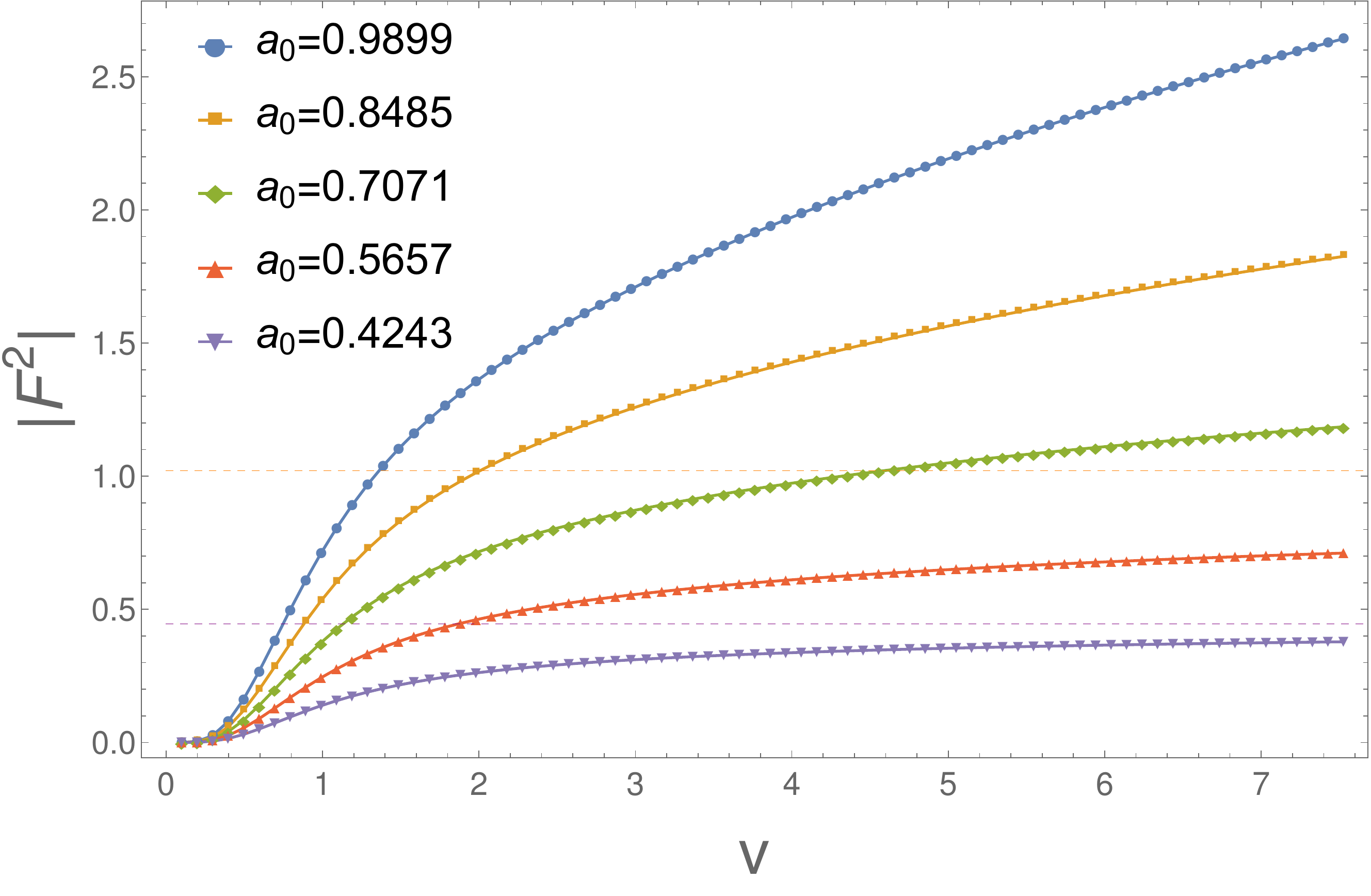}
\caption{$|F^{2}|$ at $\xi = 0, \eta = 1$ as a function of $v$ for five different values of $a_{0}.$ For the sub-critical amplitudes, the value of $|F^{2}|$ on the horizon in the corresponding stationary solution can be deduced analytically \cite{Horowitz:2014gva} and is shown as a dotted line.}
\label{f2}
\end{figure}

\begin{figure}
\centering
\includegraphics[scale=0.3]{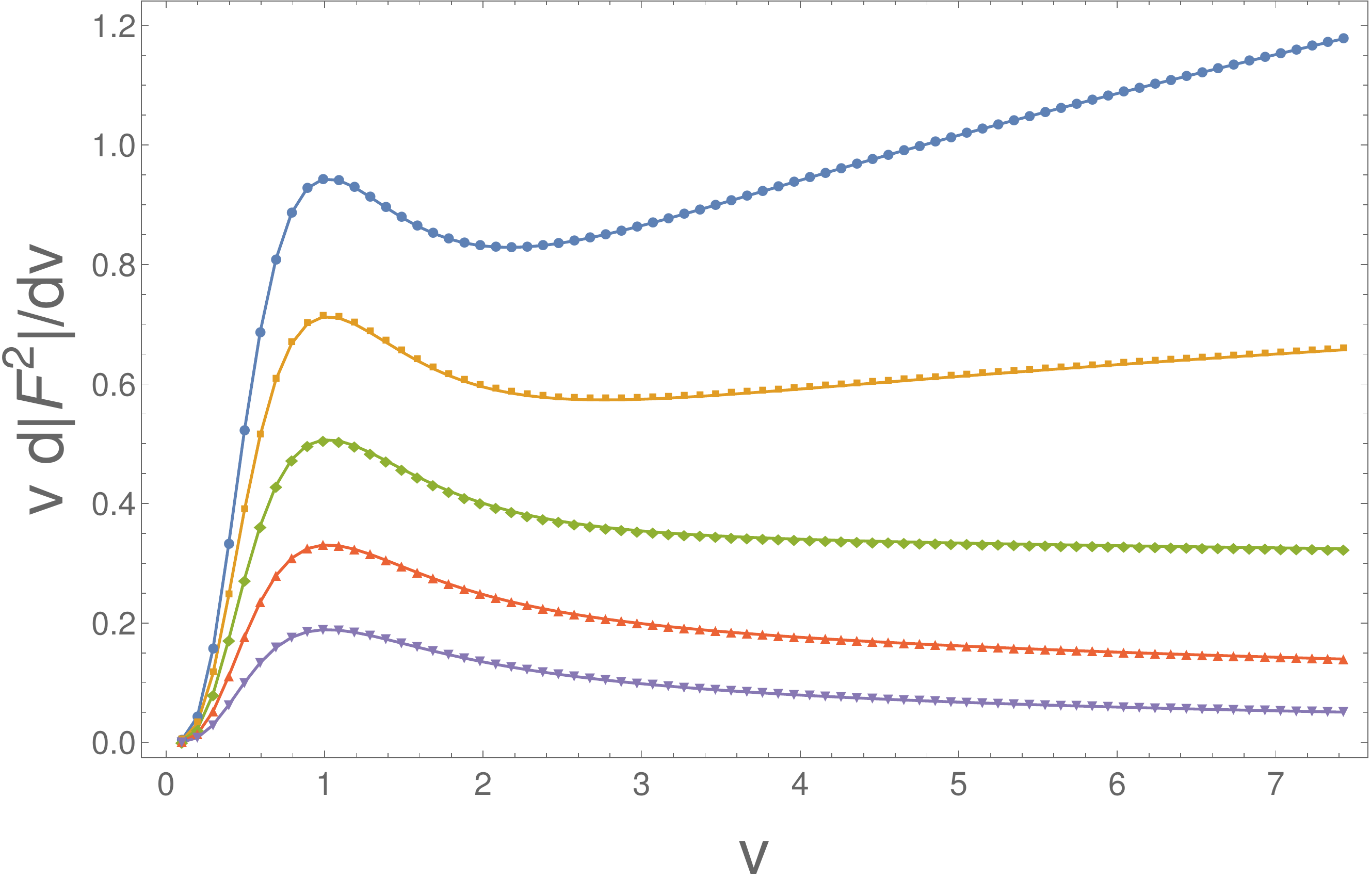}
\caption{$v \frac{d(F^2)}{dv}$ at $\xi = 0, \eta = 1$ as a function of $v$ for five different values of $a_{0}.$ The symbols and colours are the same as in Fig. \ref{f2}.}
\label{f2deriv}
\end{figure}

In Fig. \ref{f2} we plot $|F^{2}(v)|_{\xi = 0,\eta = 1}$ for our five solutions. In all five cases, the magnitude of $F^2$ is increasing with time, with decreasing derivative. The important question is whether $F^2$ will converge to some finite value, or whether its magnitude will continue to grow indefinitely. To address this we need to check whether its derivative is tending to zero faster than $1/v.$ In Fig. \ref{f2deriv} we plot $v \frac{\mathrm{d} |F^{2}|}{\mathrm{d} v}|_{\xi = 0,\eta = 1}$ for the five solutions. The solutions of sub-critical amplitude are consistent with non-divergent growth in $F^2,$ as  $v \frac{\mathrm{d} |F^{2}|}{\mathrm{d} v}|_{\xi = 0,\eta = 1}$ appears to decay as $v$ increases. The solutions of super-critical amplitude indicate a divergent growth in $F^2.$ For the solutions with the two largest amplitudes, $v \frac{\mathrm{d} |F^{2}|}{\mathrm{d} v}|_{\xi = 0,\eta = 1}$ has started to grow as $v$ increases. We also expect the $a_{0} = 0.7071$ solution to diverge, as it is super-critical, but it is difficult to deduce this with confidence from Fig. \ref{f2deriv}. We have continued to evolve this particular amplitude beyond $v = 10$ and find that, although $v \frac{\mathrm{d} |F^{2}|}{\mathrm{d} v}|_{\xi = 0,\eta = 1}$ has still not begun to grow, it does not appear to be decaying towards $0.$ This could be indicative of logarithmic growth in $F^2$ or, more likely, power law growth with a very small exponent. This can be understood from the fact that this amplitude is very close to the critical one.

\begin{figure}
\centering
\includegraphics[scale=0.3]{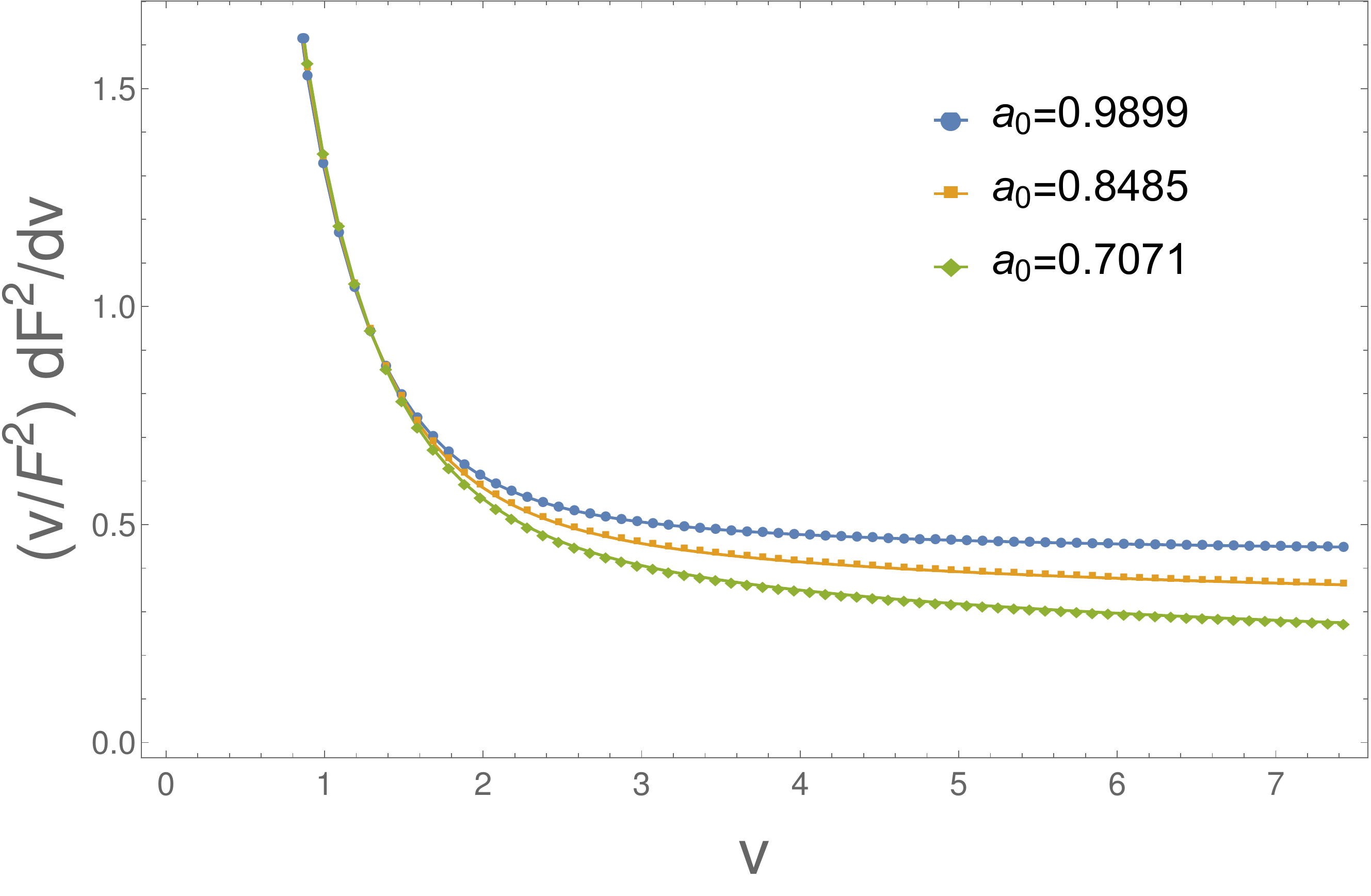}
\caption{$(v/F^2) \frac{\mathrm{d}(F^2)}{\mathrm{d}v}$ at $\xi = 0, \eta = 1$ as a function of $v$ for three different values of $a_{0}.$}
\label{f2logderiv}
\end{figure}

It is interesting to ask how fast $F^2$ is diverging in the super-critical solutions. If we assume that its late time growth is governed by a power law $F^2 \sim v^{\gamma}$ then the logarithmic derivative $\frac{v}{F^2} \frac{\mathrm{d} F^2}{\mathrm{d} v}|_{\xi = 0,\eta = 1}$ would equal the exponent $\gamma.$ In Fig. \ref{f2logderiv} we plot the logarithmic derivative of our super-critical solutions. The results are consistent with power law growth at late times, with an exponent that increases with the amplitude.

We have so far presented evidence that the curvature is growing on the \emph{apparent} horizon $\xi = 0.$ However, the apparent horizon itself will generically lie inside the event horizon and so will not be visible to distant observers. In numerical investigations of the WCCC, it is usually not possible to locate the event horizon, and instead it is assumed that singular behaviour on the chosen apparent horizon must lead to singular behaviour on the event horizon as well. We can take the same approach here, but it would be preferable to show explicitly that it is possible for a null geodesic to travel from a region of arbitrarily large curvature out to the conformal boundary. We claim that our results allow us to do this. We have evidence that the apparent horizon is approaching the true event horizon at late times.

\begin{figure}
\centering
\includegraphics[scale=0.3]{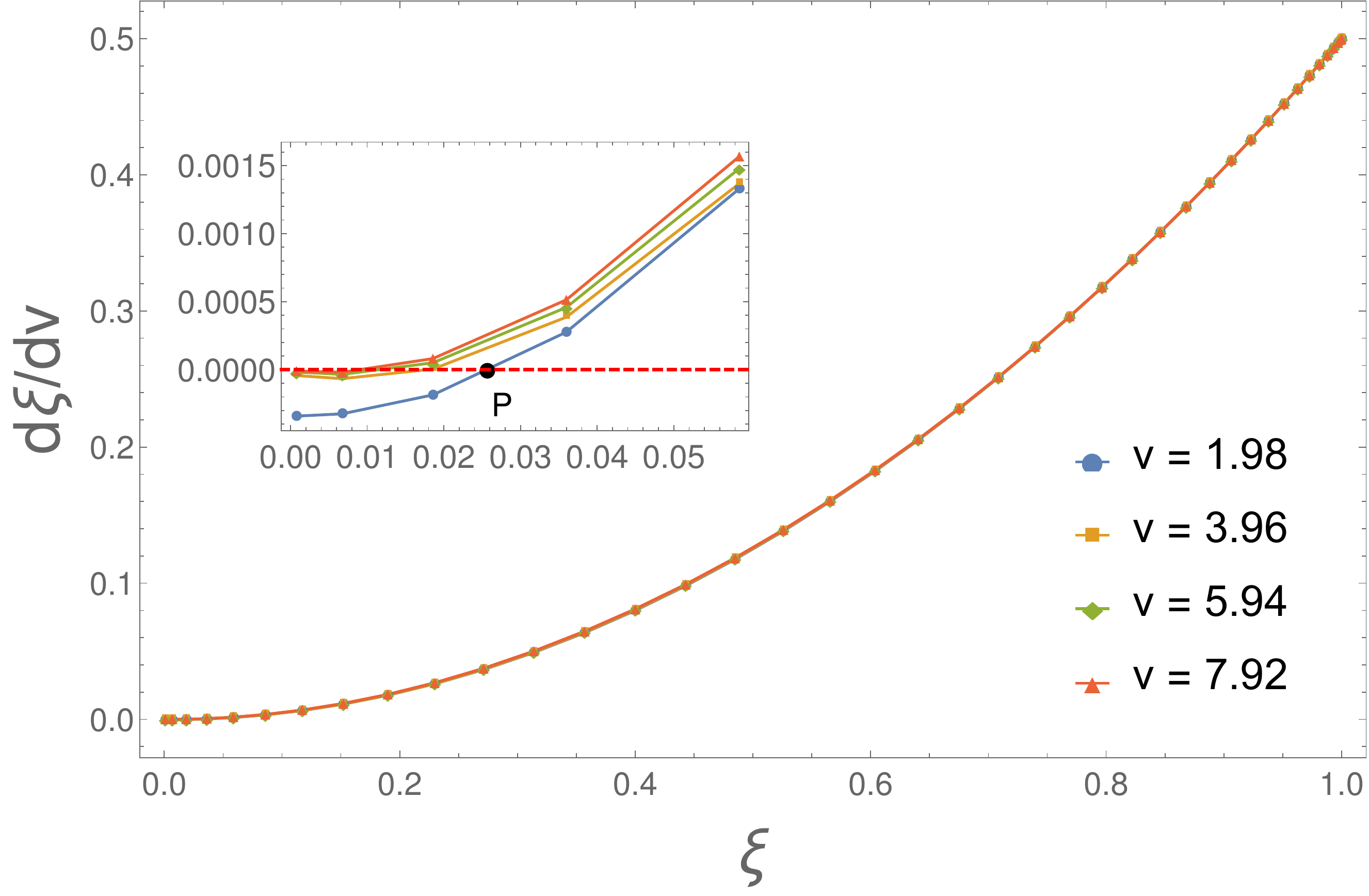}
\caption{Coordinate velocity of an outgoing null geodesic along $\eta = 1$ for the $a_{0} = 0.7071$ solution at various times. The inset plot depicts small $\xi.$}
\label{outgoingvelocityplot}
\end{figure}

In Fig. \ref{outgoingvelocityplot} we plot the coordinate velocity $\frac{\mathrm{d}\xi}{\mathrm{d}v}$ of an outgoing null geodesic along $\eta = 1$ as a function of $\xi$ for the amplitude $a = 0.7071$ solution at times $v \approx 1.98, v \approx 3.96, v \approx 5.94, v \approx 7.92.$ We find a function which is increasing and positive almost everywhere except for very close to the apparent horizon, $\xi = 0.$ If this plot were independent of time, then an outgoing null geodesic to the right of the zero (marked with the letter $P$ in Fig. \ref{outgoingvelocityplot} for the $v = 1.98$ case) would eventually reach the conformal boundary. An outgoing null geodesic to the left of the zero would fall towards $\xi = 0.$ The zero itself would mark the position of the event horizon, although the relevant quantities are so small here that we should be worried about numerical error.

\begin{figure}
\centering
\includegraphics[scale=0.3]{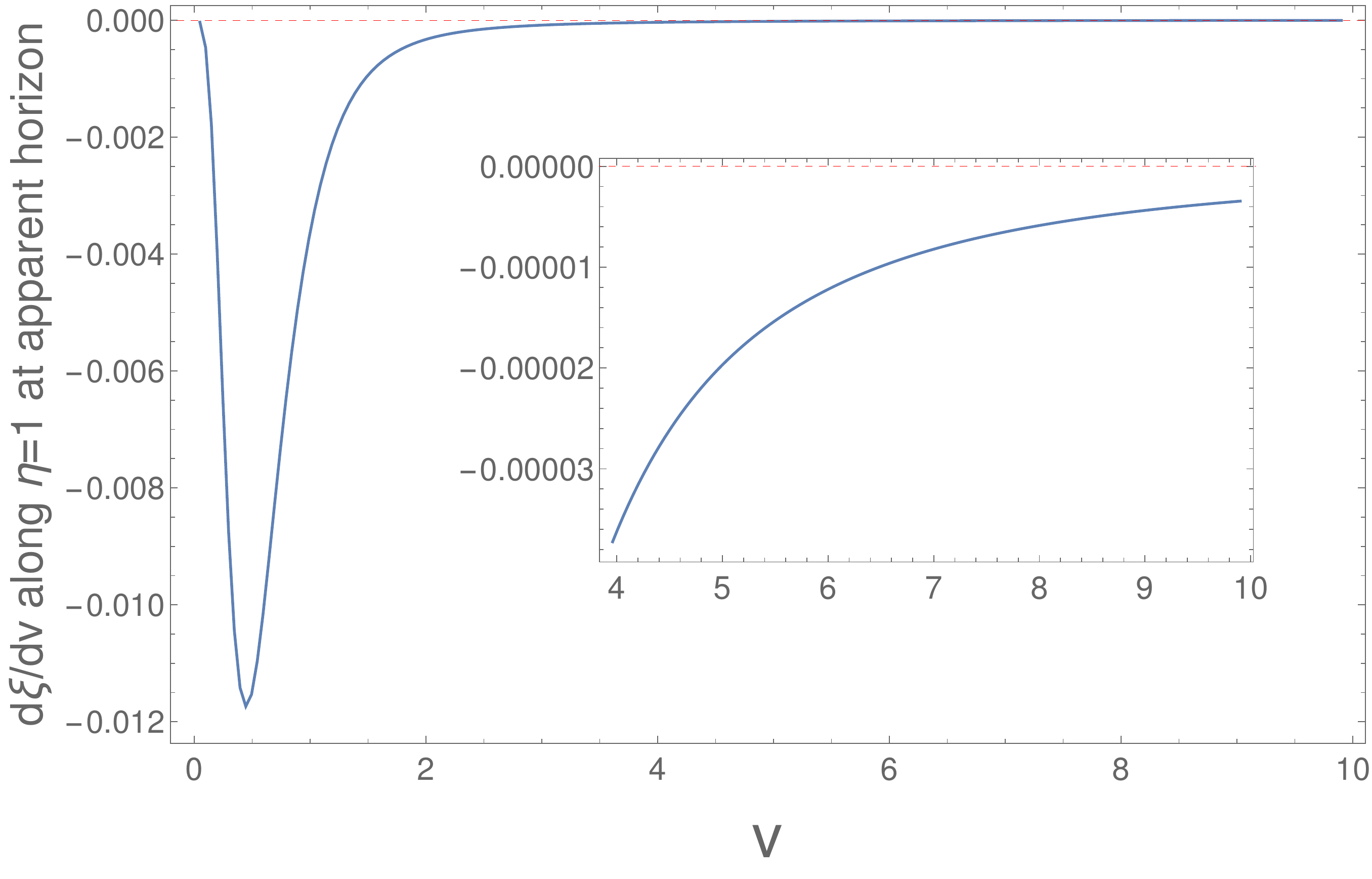}
\caption{Coordinate velocity of an outgoing null geodesic along $\eta = 1$ at $\xi = 0$ for the $a_{0} = 0.7071$ solution as a function of $v.$ The inset plot depicts late $v.$}
\label{outgoingvelonhorizon}
\end{figure}

We cannot actually identify the event horizon at any particular time because the solution \emph{is} time dependent. However, at late times the plots of Fig. \ref{outgoingvelocityplot} are only changing very slowly. They appear to be converging to some time independent function, and importantly, the zero is moving leftwards towards the apparent horizon ($\xi = 0$). This suggests that the coordinate velocity at late times is converging to a function of $\xi$ which is positive and increasing for \emph{all} $\xi > 0,$ with a zero exactly at the apparent horizon $\xi = 0.$ If this were true, then any positive $\xi$ coordinate would eventually be connected to the conformal boundary by an outgoing null geodesic and the apparent horizon would be approaching the event horizon at late times. To check this, in Fig. \ref{outgoingvelonhorizon} we plot the coordinate velocity at $\xi = 0$ itself to show that it is indeed approaching zero with increasing $v.$

A final possible concern is that even if the apparent horizon approaches the event horizon at late times, and even if the curvature on the apparent horizon is increasing without bound, the approach may be such that the curvature on the event horizon remains bounded. We discuss this possibility further in the supplemental material, and argue that it does not appear to be consistent with our results.

{\bf~Conclusions --} Our results show that the WCCC does not hold in four-dimensional AdS spacetimes whose constant time slices have planar topology. Furthermore, we have seen that our initial data far in the past is just pure AdS, and as such has no trapped surface. In the future, the curvature grows without bound leaving regions of spacetime with arbitrarily large curvatures naked to boundary observers.

One might think that, since we are increasing the electric field with time, the system will heat up according to the Joule effect and thus not be consistent with imposing zero temperature at large values of $R$ ($\xi=\eta=0$ in the coordinates used in the numerics), see for instance \cite{Horowitz:2013mia}. However, raising the temperature of a planar horizon requires an infinite amount of energy and we have checked that the energy we add to the system is finite. Our deformation is local on the boundary, in the sense that $f(t,R)$ decreases at large $R$ as $1/R^\alpha$ with $\alpha\geq1$. For such decays, the dominant contribution to the holographic stress energy tensor $\langle T_{tt}\rangle$ comes from normalisable spherical waves, \emph{i.e.} quasinormal modes of the system, which contribute as $\langle T_{tt}\rangle \sim R^{-1}$. One might worry that such contributions will make the total energy of the system infinite, and if we integrate $\langle T_{tt}\rangle$ over a constant $v$ slice of the boundary, it does indeed diverge. However, we find that this $1/R$ tail in $\langle T_{tt}\rangle$ decays exponentially with $v$ so that when we integrate over a constant $t$ slice the result is finite and bounded as $t \rightarrow \infty.$ This implies that the total energy added to the system is finite.

Our counterexample to the WCCC is likely to become more intricate if electrically charged particles with charge $q$ and mass $m$ are included into our action (\ref{eq:action}). We expect that if $|q|\ll m$, the WCCC is still likely to be violated. The most natural scenario is that a critical value for $|q|/m\sim 1$ exists above which the charged hovering black holes of \cite{Horowitz:2014gva} can form. This raises the intriguing possibility of link between the WCCC and the weak gravity conjecture \cite{ArkaniHamed:2006dz}. Whether or not black holes will form precisely for $|q|/m\geq1$, which is the inequality predicted by the weak gravity conjecture, is something we hope to address in the near future. Even if charged matter exists in the theory, classically one can set it to zero and it will not affect the solutions discussed in this paper, though such initial data would look fine tuned. However, even in this case, quantum particle creation might still weaken the electric field on the horizon \footnote{We thank G.~Horowitz for a discussion about such initial data.}.

One might wonder if our initial data is generic enough, since $\partial/\partial \phi$ is  a Killing vector field \emph{everywhere} in the bulk. However, it is clear that considering initial data with no such restriction is unlikely to change our conclusions. The reason being that even if $\partial/\partial \phi$ is only an asymptotically Killing vector field, no static solutions with a simply connected horizon were found in \cite{Horowitz:2014gva,Kunduri:2013ana} for $a>a_{\star}$.

Finally, we have explicitly checked that, even with the time dependent boundary conditions imposed in this paper, the metric and gauge field satisfy the required conditions for the proof of the positivity of energy theorem detailed in \cite{Gibbons:1983aq}. This result does not follow immediately from the coordinates used in the numerics. However, one can show that if we transform our boundary expansion to Fefferman-Graham coordinates \cite{Fefferman:2007rka}, the approach to pure AdS in standard Poincar\'e coordinates is compatible with those required in \cite{Gibbons:1983aq}.

{\bf~Acknowledgements --} TC is supported by an STFC studentship. We would like to thank \'O.~J.~C.~Dias for reading an earlier version of this manuscript and providing many insightful comments. We would like to thank Gary Horowitz and Harvey Reall for numerous helpful discussions.

\newpage
\onecolumngrid
\appendix

\section{Supplemental material}

\subsection{Details of the numerical method}

To implement our numerics we define coordinates $y$ and $r$ which are related to $\xi$ and $\eta$ by\begin{align}
\xi &= \eta \left(1 + \frac{1 - y^2}{2 y^2}\right)^{-1}\\
\eta &= \frac{1 - r^2}{\sqrt{r^2 + (1-r^2)^2}}.
\end{align} Both $y$ and $r$ take values between $0$ and $1.$ $y = 0$ is the apparent horizon, $y = 1$ is the conformal boundary, and $r = 0$ is the symmetry axis.

We use spectral methods to integrate our equations along the characteristic curves (curves of constant $r$ and $v$). All
of our functions can be extended smoothly through $r = 0$ to be invariant under
$r \rightarrow -r.$ This follows from the $SO(2)$ symmetry. We therefore use a Chebyshev grid for the doubled domain $-1 < r < 1$ and construct differentiation matrices to make
this symmetry explicit. We have also defined $y$ so that we are able to
extend our functions through $y = 0$ in the same way. This works as long as the
functions have a power series expansion in $\xi$ in the neighbourhood of $\xi = 0.$
In the new coordinates this becomes a power series in $y^2.$ We therefore use
the same doubling trick for $y,$ choosing a Chebyshev grid on $-1 < y < 1.$ There
is then no point on the horizon itself ($y = 0$) which simplifies the evaluation
of our equations. We use a grid of 30 points in the $r$ direction and 40 points in the $y$ direction. In order to evolve forward to the next null slice we use the fourth order Runge-Kutta method. To obtain stable solutions, we find it necessary to apply filtering in both the $r$ and $y$ directions. We do this every few time steps by interpolating onto a grid with $2/3$ the number of points and then back to the original grid again.

We now present a full list of the functions we solve for (denoted with a subscript $n$), the boundary conditions we apply, and the order in which we tackle the equations. This closely follows the method described in \cite{Balasubramanian:2013yqa} but we also need to include a Maxwell field. First, as described in the main text, we partially fix the gauge by choosing a $\xi$ dependence for $\chi$\begin{equation}
\chi(v,\xi,\eta) = \chi_{n}(v,\eta) \frac{(\eta - \xi)^3}{\eta^3}.
\end{equation}
It is also useful to define the new functions
\begin{subequations}
\begin{align}
\pi_{1} &= \eta e^{2\chi - 2\beta} \left(F_{v \xi} + U F_{\xi \eta}\right)\\
\pi_{2} &= - \eta^3 \frac{1}{(\eta - \xi)^2} e^{2\chi - 2\beta + \alpha} e^{6 [-\frac{1}{3}(\xi/\eta)^3 + (\xi/\eta)^2 -  (\xi/\eta)] \chi_{n}} \xi^2 \frac{\mathrm{d} U}{\mathrm{d} \xi}.
\end{align}
\end{subequations}
The remaining functions we solve for are defined via:
\begin{subequations}
\begin{align}
\beta &= \frac{1}{8} (2 - y^2)^3 (1 - y^2)^3 \left(\beta_{n}-\chi_{n}\right)
\\
U &= \frac{1}{4} (2 - y^2)^2 (1 - y^2)^2 \frac{r^2}{1-r^2+r^4} U_{n}
\\
F_{v \xi} &= \frac{y^4 (2-y^2)^2 (1-r^2)}{\sqrt{1-r^2+r^4}} (F_{v \xi})_{n}
\\
V &= 1 - \frac{(1-y^2)^4 (1+y^2) \sqrt{1-r^2+r^4}}{2 y^4 (1-r^2) \left[(1+y^2)^3 - 3(1-y^2)^3 \chi_{n} \right]} \left[\chi_{n} + \frac{1+y^2}{1-y^2} e^{\frac{4 y^2 (3+y^4) \chi_{n}}{(1+y^2)^3}} V_{n}\right]
\\
\alpha &= -\log\left(\frac{r^2}{1-r^2+r^4}\right)+ \frac{1}{4} (2 - y^2)^2 (1 - y^2)^2 r^2 \alpha_{n}
\\
F_{\xi \eta} &= y^4 (2-y^2)^2 (F_{\xi \eta})_{n}
\\
F_{v \eta} &= \frac{\sqrt{1-r^2+r^4}}{1-r^2} (F_{v \eta})_{n} + \frac{1}{2} F_{\xi \eta}(V - 1)
\\
\frac{\mathrm{d}\alpha}{\mathrm{d}v} &= \frac{1}{4} (2 - y^2)^2 (1 - y^2)^2 r^2 \dot{\alpha}_{n} - \frac{1}{2} \xi^2 \frac{\mathrm{d} \alpha}{\mathrm{d} \xi} (V - 1)
\\
\frac{\mathrm{d}\chi}{\mathrm{d}v} &= \dot{\chi}_{n}
\\
\frac{\mathrm{d}(F_{\xi \eta})}{\mathrm{d} v} &=  y^4 (2-y^2)^2 (\dot{F}_{\xi \eta})_{n}.
\end{align}
\end{subequations}

Now given $\alpha_{n},\ (F_{\xi \eta})_{n},\ \chi_{n}$ and Dirichlet data for $V_{n},\ \pi_1,\ \pi_2$ on an initial null slice, we solve Einstein's equations $E_{a b} = G_{a b} + \Lambda g_{a b} - 8 \pi T_{a b} = 0$ and Maxwell's equations as follows:\begin{enumerate}
\item Solve for $\beta_n$ using the $\xi \xi$ component of Einstein's equations, imposing $\beta_{n}|_{y = 1} = 0$ as a boundary condition.
\item Solve for $\pi_1$ using the $v$ component of Maxwell's equations, imposing the given Dirichlet boundary condition for $\pi_1.$
\item Solve for $\pi_2$ using the $\xi \eta$ component of Einstein's equation, imposing the given Dirichlet boundary condition for $\pi_2.$
\item Solve for $U_n$ by inverting the definition of $\pi_2$ and imposing $U_n|_{y=1} = 0$ as a boundary condition.
\item Solve for $(F_{v \xi})_{n}$ by inverting the definition of $\pi_1.$
\item Solve for $V_{n}$ using the combination $e^{-\alpha} E_{\eta \eta} + e^{\alpha} E_{\phi \phi},$ imposing the given Dirichlet boundary condition for $V_{n}.$
\item Imposing that the expansion at $y = 0$ does not change in time (so that $y = 0$ remains an apparent horizon) gives $\frac{\mathrm{d} V_{n}}{\mathrm{d} v}$ at $y = 0.$ Solve for $\dot{\chi}_{n}$ using this and the combination $E^{\xi}_{v}+U E^{\xi}_{\eta}|_{y = 0},$
 imposing $\dot{\chi}_{n}|_{r = 1} = 0$ and $\partial_r \dot{\chi}_{n}|_{r = 0} = 0$ as boundary conditions.
 \item Solve for $\dot{\alpha}_{n}$ and $(F_{v \eta})_{n}$ using the $\phi \phi$ component of Einstein's equations and the $\eta$ component of Maxwell's equations, imposing $\dot{\alpha}_{n}|_{y = 1} = 0$ and $(F_{v \eta})_{n} = (F_{v \xi})_{n} + \eta a(v)$ as boundary conditions.
 \item Solve for $(\dot{F}_{\xi \eta})_{n}$ using $\mathrm{d} F = 0.$
 \item Solve for $\frac{\mathrm{d}\pi_{1}}{\mathrm{d} v}|_{y = 1}$ using the $\xi$ component of Maxwell's equations on the boundary.
 \item Solve for $\frac{\mathrm{d}\pi_{2}}{\mathrm{d} v}|_{y = 1}$ using the $v \eta$ component of Einstein's equations on the boundary.
 \item Solve for $\frac{\mathrm{d} V_{n}}{\mathrm{d} v}|_{y = 1}$ using the $v v$ component of Einstein's equations on the boundary.
 \end{enumerate}
 
 We now have sufficient information to obtain $\alpha_{n},\ (F_{\xi \eta})_{n},\ \chi_{n}$ and Dirichlet data for $V_{n},\ \pi_1,\ \pi_2$ on the next null slice, using the fourth order Runge-Kutta method. We then repeat the procedure.
 
\begin{figure}
\centering
\includegraphics[scale=0.3]{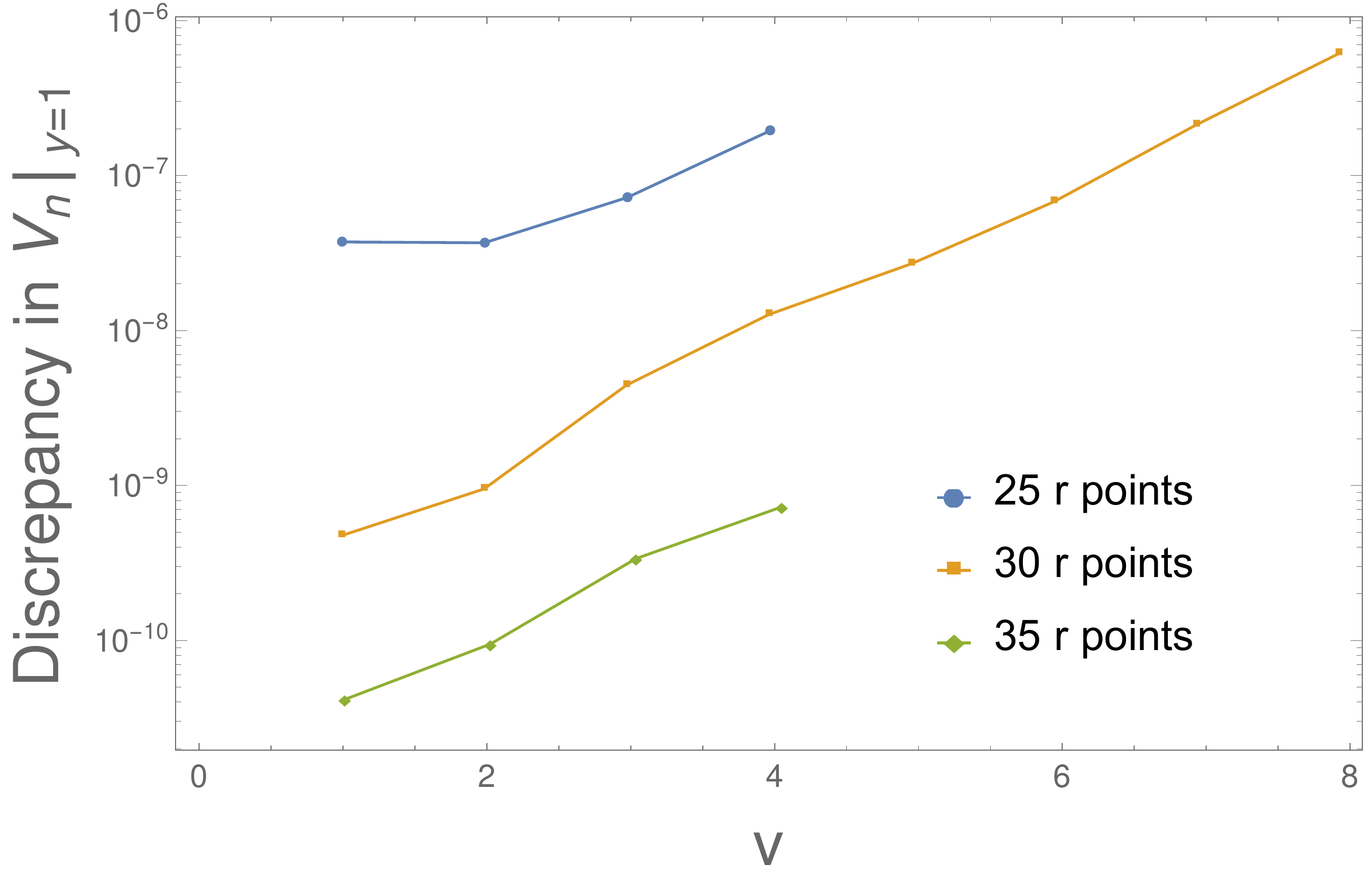}
\caption{A log plot of the maximum difference between the two methods of calculating $V_{n}|_{y = 1}$ for the $a_{0} = 0.7071$ solution. We vary the number of points in the $r$ direction and use 40 points in the $y$ direction.}
\label{constraintviolation}
\end{figure}

To check the accuracy of our numerics, we can compare two different ways of computing $V_{n}|_{y = 1}.$ In our numerical scheme, we chose to evaluate $V_{n}|_{y=1}$ by using the boundary constraint equation $E_{v v}$ at $y = 1$ to evolve forward from the previous null slice. Alternatively, we could have imposed zero expansion at $y = 0$ and used this to obtain a Dirichlet boundary condition for $V_{n}$ at $y = 0$ instead, which gives us a different way of performing step 6. This would allow us to determine $V_{n}$ everywhere without assuming knowledge of $V_{n}|_{y = 1}.$ If we had done this, the $E_{v v}$ constraint would instead be imposed at the horizon through step 7. We should be free to impose the boundary constraint at either the boundary or the horizon, but we expect the two approaches to disagree due to numerical error. Comparing the alternative ways of calculating $V_{n}|_{y=1}$ therefore gives a non-trivial check on numerical accuracy. In Fig. \ref{constraintviolation}, we plot the maximum size of the difference between the two methods of calculating $V_{n}|_{y = 1}$ against $v$ for our $a_{0} = 0.7071$ solution. As a test on convergence, we have repeated this for different grid sizes.

\begin{figure}
\centering
\includegraphics[scale=0.3]{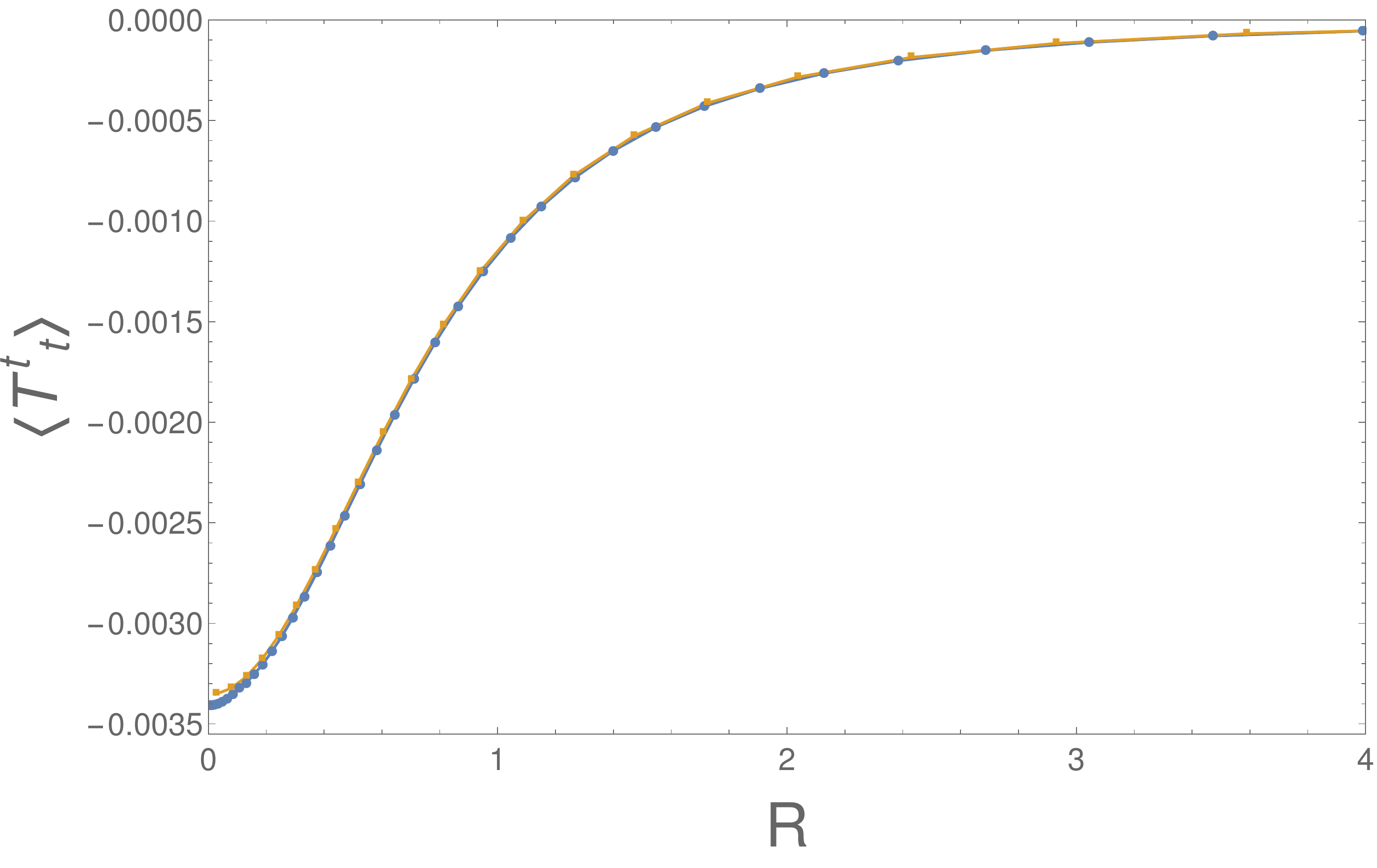}
\caption{Comparing the numerical results of \cite{Horowitz:2014gva} (blue disks) and the current simulations (orange squares) for $a_0 = 0.4243$.}
\label{comparison}
\end{figure}

As a final check on our numerics, we can compare the end-point of our sub-critical solutions to the stationary solutions obtained in \cite{Horowitz:2014gva}. We make this comparison in Fig. \ref{comparison}, plotting the $t t$ component of the boundary stress tensor in the two cases, and finding good agreement. The small discrepancy can be explained by the fact that our solution has not yet completely settled down.

\subsection{Does diverging curvature on the apparent horizon imply diverging curvature on the event horizon?}

In the main text, we presented evidence that the curvature was growing without bound on the apparent horizon, and that the apparent horizon was approaching the event horizon at late times. We would like to conclude that the curvature is growing without bound on the event horizon as well. Here we attempt to make this argument more precise.

One way to argue that the curvature is diverging on the event horizon is as follows. Given an arbitrarily large bound $C:$\begin{itemize}
\item after some finite time, by extrapolating our numerical results, we will have $R_{a b} R^{a b} > C$ on the apparent horizon $\xi = 0.$
\item By continuity, at the same time there will be a point with small positive $\xi$ coordinate, $\xi = \delta,$ where the curvature also violates this bound.
\item After some additional finite time, by extrapolating our numerical results again, the point with $\xi = \delta$ will be visible to boundary observers.
\item If we additionally assume that the curvature has not decreased at $\xi = \delta$ in this time, then we now have a point with $R_{a b} R^{a b} > C$ visible to boundary observers.
\end{itemize}

\begin{figure}
\centering
\includegraphics[scale=0.3]{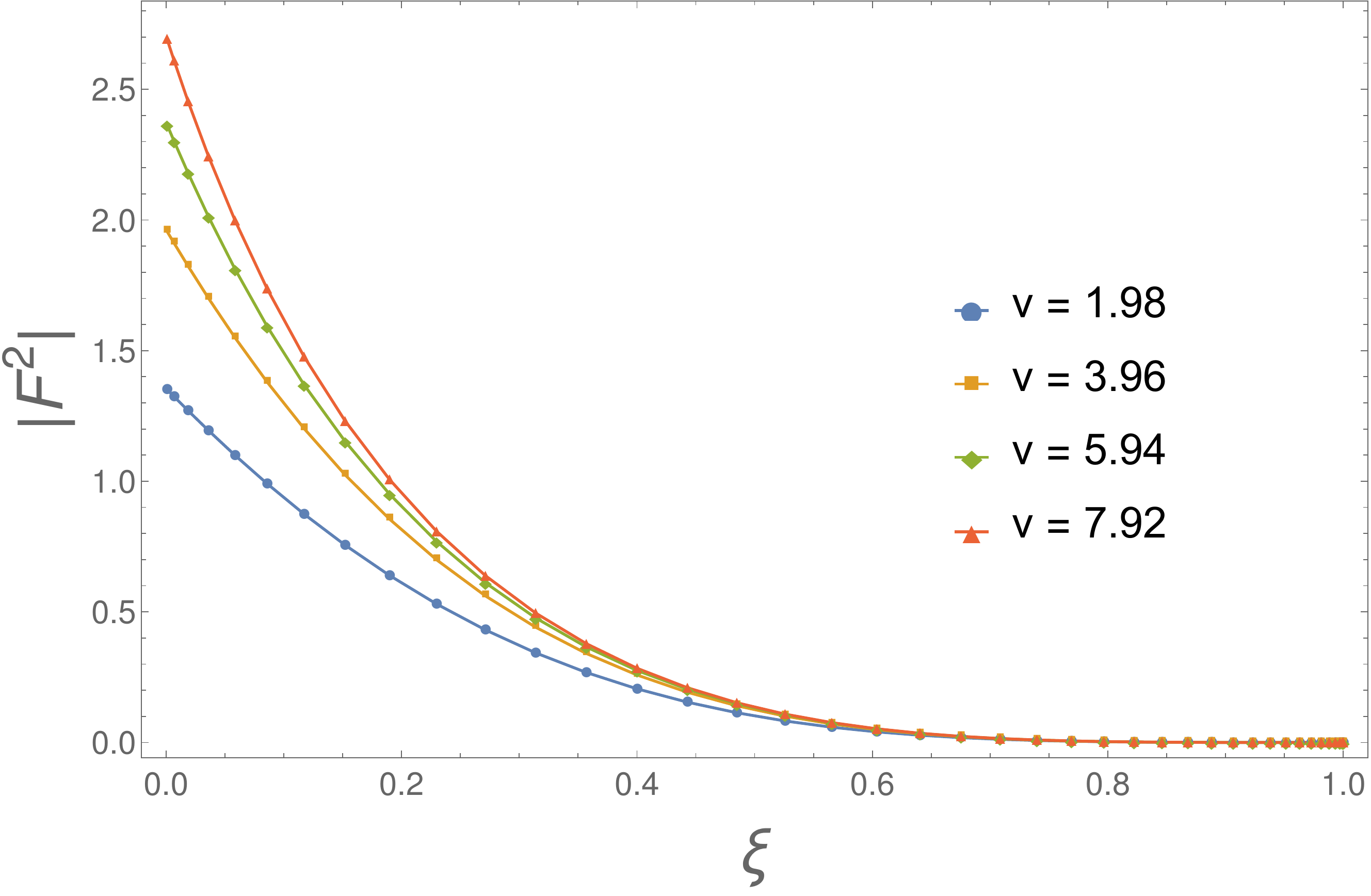}
\caption{Plot of $|F^2|$ against $\xi$ along $\eta = 1$ for the $a_{0} = 0.9899$ solution at various times.}
\label{increasing}
\end{figure}

The assumption that $R_{a b} R^{a b}$ is always increasing with $v$ for fixed $(\xi,\eta)$ was essential to this argument, and this does turn out to be the case, at least over the range of $v$ values for which we have numerical results. To show this, in figure \ref{increasing}, $F^2$ is plotted along $\eta = 1$ for the $a_{0} = 0.9899$ solution at various times.

\bibliography{refs}{}
\end{document}